\documentclass[12pt]{article}  
\usepackage{a4,epsfig}
\begin{document}


\newcommand {\eeGg} {$e^+e^- \rightarrow G\gamma$ \hspace*{1mm}}
\newcommand {\eenngg} {$e^+e^- \rightarrow \nu\bar{\nu}\gamma(\gamma)$
  \hspace*{1mm}}
\newcommand {\eeeeg} {$e^+e^- \rightarrow e^+e^-\gamma$ \hspace*{1mm}}



\begin{titlepage}

\pagenumbering{arabic}
\vspace*{-1.5cm}
\begin{tabular*}{15.cm}{lc@{\extracolsep{\fill}}r}
{\bf DELPHI Collaboration} & 
\hspace*{1.3cm} \epsfig{figure=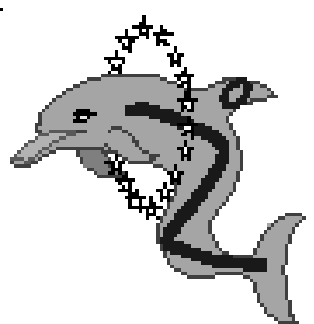,width=1.2cm,height=1.2cm}
&
DELPHI 2007-001 CONF 760
\\
& &
5 July, 2007
\\
&&\\ \hline
\end{tabular*}
\vspace*{2.cm}
\begin{center}
\Large 
{\bf \boldmath

Search for one large extra dimension with the DELPHI detector at LEP2 

} 
\vspace*{2.cm}
\normalsize { 
   {\bf S. Ask, V. Hedberg and F.-L. Navarria}\\
   {\footnotesize CERN, Geneva, Switzerland}\\ 
   {\footnotesize Dept. of Physics, Lund, Sweden}\\ 
   {\footnotesize Dip. di Fisica and INFN, Bologna, Italy}\\
   
}
\end{center}
\vspace{\fill}
\begin{abstract}
\noindent

Single photons detected by the DELPHI experiment at LEP2 in the years
1997-2000 are used to investigate the existence of a single extra
dimension in a modified ADD scenario with slightly warped large extra 
dimensions.
The data collected at centre-of-mass energies between 180 and 209 GeV
for an integrated luminosity of $\sim$ 650 $\mathrm {pb^{-1}}$ agree with the
predictions of the Standard Model and allow a limit to be set 
on graviton emission in one large extra dimension.
The limit obtained on the fundamental mass scale $M_D$ is 1.69 TeV
at 95$\%$ CL, with an expected limit of 1.71 TeV.



\end{abstract}
\vspace{\fill}
\begin{center}
Contributed Paper for Lepton Photon 2007 (Daegu, Korea)
\end{center}
\vspace{\fill}
\end{titlepage}







\section{Introduction}

The Standard Model (SM) has been thoroughly tested at the CERN LEP $e^+e^-$
collider~\cite{lep}. No sign of statistically significant deviations from 
it or evidence for new physics phenomena beyond it have been found up to 
the highest LEP centre-of-mass energies of about 209 GeV. 
Yet the SM cannot be the final picture, because of several 
theoretical problems. One is known as the hierarchy 
problem and is related to the observed weakness of gravity in comparison 
with other interactions. This may be expressed by the observation that 
the reduced Planck mass, 
$M_{Pl} = \sqrt{1/G_N} \sim 2.4 \cdot 10^{15}$ TeV, where $G_N$ is  
Newton's coupling constant, is much larger than the 0.1-1 TeV scale 
of the electroweak symmetry breaking. 
\par
A step towards the 
solution of this puzzle was proposed in 1998 by Arkani-Hamed, 
Dimopoulos and Dvali (ADD)~\cite{add}, assuming the existence of 
large extra 
spatial dimensions (ED). One ED was proposed a long time ago in 
connection with gravity and its unification with electromagnetism in the 
papers of Kaluza and Klein (KK)~\cite{kaluza}. 
More recently, with the appearence of 
string theory, several EDs were advocated, but their size 
was thought to be close to the Planck length, $R \sim 1/M_{Pl} 
\sim10^{-33}$ cm. In this case EDs would be completely out of the reach of 
present and 
planned colliders. 
The 
novel 
suggestion of ADD 
was the possible existence of large EDs with a fundamental Planck mass close 
to the electroweak scale, in fact implying that non-trivial physics ``ends'' 
at energies of about 1 TeV. In the ADD model all the SM particles are 
supposed
to live on a 3D brane corresponding to our usual 
space, while gravitons are 
allowed to propagate into the bulk. Thus the weakness of gravity is 
simply due to its dilution in the volume of the EDs. 
\par  
Assuming flat EDs and compactification on a torus, Gauss' law gives: 

\begin{equation}
M^2_{Pl} = R^n M_D^{n+2},
\label{eq:gausslaw}
\end{equation}

\noindent where $R$ is the radius of the ED and $M_D$ is the fundamental Planck 
scale in the D-dimensional space-time (D=4+$n$). With $M_D \sim$1 TeV 
and $n$=1, eq.~\ref{eq:gausslaw} implies a modification of Newton's law over solar system 
distances which is not observed. 
So the possibility that $n$=1 
is usually considered to be falsified. 
On the other hand for $n\geq$2, $R<$1~mm and tests of gravity are 
only recently reaching these small distances~\cite{coyle}. For $n\geq$3, 
$R<$1~nm and no gravity test exists which can falsify the model. 
\par
The graviton, confined within flat EDs of size $R$, has a uniform spectrum 
of excitations, which, from the point of view of a 4D observer, will 
be seen as a KK tower of states, with masses uniformly spaced between 
$1/R$ $(\sim 10^{-32/n}$ TeV) and $M_D$. In particle
collisions at accelerators and in the cosmos, gravitons can be emitted,  
but they escape immediately into the bulk, with momentum conservation in all the dimensions, and 
are therefore 
detectable via a missing energy signature. 
Each KK state is very weakly coupled, yet the number of states 
is very large, which turns into a sizable cross-section for graviton 
emission. Astrophysics yields strong constraints for $n$=2,3 
based on observations of supernova SN1987A 
and on the behaviour 
of neutron stars~\cite{astro}. The limits vary from 20 to 40 TeV and 2 to 3 TeV, 
respectively, 
and seem to rule out the ADD model with $M_D$=1 TeV. They are 
however based on many assumptions with differences of a factor of 2-3 
between different calculations. 
For larger $n$ they become 
much weaker. 
\par 
For $n\geq$2 limits on graviton emission have been obtained at the 
LEP collider~\cite{{al},{de},{l3},{op}} and at the Tevatron~\cite{tevatron}. 
At LEP the direct graviton emission reaction \eeGg ($GZ$) has been 
studied: for $n\geq$2 the photon spectrum peaks at low energies 
and at small emission angles~\cite{grw}. No excess with respect to the SM 
predictions has been found and a combination of the LEP results yielded  
$M_D>$1.60 (0.80) TeV for $n$=2 (6) at the 95$\%$ Confidence Level (CL)~\cite{wg}.  
\par 
Recently the ADD model has been reconsidered by Giudice, Plehn and 
Strumia (GPS)~\cite{gps}, who have focused on the infrared (IR) behaviour of the model in 
connection with limits at colliders versus gravity and astrophysics 
constraints. 
They considered a distorted version of the ADD model with the same 
properties in the ultraviolet (UV) region, but satisfying observational and astrophysical 
limits in the large distance regime. 
They showed that the introduction of an IR cut-off in 
the ADD model evades the constraints from astrophysics and gravity 
for small $n$, including  
$n$=1, given the energy resolution of the collider experiments. 
This IR cut-off is equivalent to a slight deformation or warping of the 
otherwise flat EDs. They started from the Randall and Sundrum type 1 
model (RS1)~\cite{rs1} and considered the limit 
of slightly warped but large ED, resulting in a moderately large 
total warp factor.
In RS1, with the visible brane located at $y$=0 and the 
Planck brane at $y=\pi R$, the line element is non-factorizable due 
to the warping factor 
\begin{eqnarray}
ds^2 = e^{2 \sigma(y)} \eta _{\mu \nu} dx^{\mu}dx^{\nu} + dy^2 
\end{eqnarray}
\noindent with $\sigma(y) = \mu |y|$. Here $\mu$ is a mass parameter due to the warp and $y$ is the coordinate 
in the extra dimension. The mass parameter
has a value 50 MeV $\leq \mu <<$ 1 TeV which introduces an IR cut-off. 
This cut-off implies a mass of the graviton which is inaccessible
for cosmological processes, but which has no significant implications 
for the high energy collider signal in the UV region of the KK spectra.
In particular, the relation between the fundamental mass 
scale in 5 dimensions 
and the 4D Planck 
mass becomes
\begin{equation}
M^2_{Pl} = \frac{M^3_5}{2 \mu \pi} \left( e^{2 \mu R \pi} - 1 \right),
\end{equation}
where $R$ is the radius of the compactified ED. 
Hence the one ED can still be large, but unobserved as a modification 
of Newton's law or in the 
cosmological low energy processes. In this model the hierarchy between 
the Fermi and Planck scales is generated by two factors, the large ED 
and warping. It can be seen that for $\mu << R^{-1}$ one obtains the 
ADD limit, eq.~(\ref{eq:gausslaw}).  
\par
Since a search for graviton emission with $n$=1 
was not performed in a previous publication~\cite{de}
and since the 
results cannot be inferred from the limits already given for $n\geq$2 
owing to the totally different photon energy spectrum~\cite{{grw},{gps}}, 
the DELPHI data were reanalysed and the results will be presented here. 
The paper is organized as follows: Section 2 recalls briefly the 
experimental details, the analysis is discussed in Section 3,  
Section 4 presents the results and the conclusions are given in 
Section 5.

\section{Detector and data preselection}

\par The general criteria for the selection of single-photon events 
are based mainly on the electromagnetic calorimeters and on the 
tracking system of the DELPHI detector~\cite{delphi}. All the three 
major electromagnetic calorimeters in DELPHI, the High density 
Projection Chamber (HPC), the Forward ElectroMagnetic Calorimeter (FEMC)
and the Small angle TIle Calorimeter (STIC), have been used in the 
single-photon reconstruction. The STIC accepted photons at very small 
polar angle~\footnote{In the DELPHI coordinate system, the z axis is 
along the electron beam direction and the polar angle to the z axis is called
$\theta$.}, 
2$^\circ~<\theta~<~10^\circ$ 
(170$^\circ~<\theta~<~178^\circ$), the FEMC covered intermediate 
angles, 10$^\circ$~$< \theta <$~37$^\circ$
(143$^\circ$~$< \theta <$~170$^\circ$), and large angles, 
40$^\circ$~$< \theta <$~140$^\circ$, were covered by the HPC.
Hermeticity Taggers were used to ensure detector hermeticity 
for additional neutral particles in the angular region around 45$^{\circ}$ 
between HPC and FEMC, not covered by the calorimeters. 
The DELPHI tracking system and the taggers were used as a veto. 
A detailed description of the trigger conditions and efficiencies of 
the calorimeters is given in a previous publication~\cite{de}, 
where the rejection of events in which charged particles were produced 
is also discussed. 
\par
The study was done with data taken during the 1997-2000 runs at $e^+e^-$ 
centre-of-mass energies from 180 to 209 GeV, corresponding to an integrated
luminosity of $\sim$ 650 $\mathrm {pb^{-1}}$, with the subdetectors relevant
for the analysis all fully operational.
\par
The single-photon events were selected in two stages. In the first stage 
events with only one detected photon were preselected and compared 
to the SM process $e^+e^-~\rightarrow~\nu\bar{\nu}\gamma$. 
A likelihood ratio method was then used to maximize the sensitivity 
in the search for graviton production with $n$=1.
\par
Events with a photon in the HPC were selected by requiring a shower 
having a scaled 
energy $x_\gamma = E_\gamma/E_{beam} >$0.06, 
$\theta$ 
between 45$^\circ$ and 135$^\circ$, and no charged particle tracks. 
Photons in the FEMC were required to have a scaled energy $x_\gamma>$0.10 
and a polar angle in the intervals 
12$^\circ$~$< \theta <$~32$^\circ$ (148$^\circ$~$< \theta <$~168$^\circ$). 
Single photons in the STIC were preselected by requiring one shower 
with a scaled energy $x_\gamma>$0.30 and with 
3.8$^\circ$~$< \theta <$~8$^\circ$ (172$^\circ$~$< \theta <$~176.2$^\circ$).
Additional details about the preselection are given in~\cite{de}. 
In the single-photon event preselection 
events with 
more than one photon were accepted
only if the other photons were at low 
angle ($\theta_\gamma$ $<$ 2.2$^\circ$), low energy ($E_\gamma<$0.8 
GeV) or within 3$^\circ$, 15$^\circ$, 20$^\circ$ from the highest 
energy photon in the STIC, FEMC and HPC respectively. 

\section{Single-photon analysis}

The single-photon analysis has been discussed in detail in~\cite{de}, 
here we will recall the main points and underline the differences 
in the present analysis. 
\par 
Single-photon events can be faked by the QED reaction 
\eeeeg if the two electrons escape 
undetected along the beampipe or if the electrons are in the detector 
acceptance but are not detected by the experiment. This process has 
a very high cross-section, decreasing rapidly with increasing energy and 
polar angle of the photon. Its behaviour together with the rapid 
variation of efficiencies at low photon energy motivates the 
different calorimeter energy cuts in the preselection and additional 
energy-dependent cuts on the polar angle in the FEMC and STIC.
\par The remaining background from the \eeeeg
process was 
calculated with the Monte Carlo program TEEG by D. Karlen~\cite{karlen} and two 
different event topologies were found to contribute, giving 
background at low and high photon energy respectively. Either both 
electrons were below the STIC acceptance or one of the electrons 
was in the DELPHI acceptance where it was wrongly identified as a photon, 
while the photon was lost for example in the gaps between the electromagnetic 
calorimeters not covered by the Hermeticity Taggers, or in masked crystals 
in the FEMC. 
\par The contribution from other processes has also been calculated: cosmic 
ray events, $\gamma\gamma$ collisions using PYTHIA 6.1 and 
DBK~\cite{gammagamma}, $e^+e^-~\rightarrow~\gamma\gamma(\gamma)$ with a 
program by Berends and Kleiss~\cite{berends1}, 
$e^+e^-~\rightarrow~\mu\mu(\gamma)$ and 
$e^+e^-~\rightarrow~\tau\tau(\gamma)$ with KORALZ~\cite{koralz}, and 
four-fermion events with EXCALIBUR and Grc4f~\cite{berends2}.
\par The \eenngg~process was simulated by the 
KORALZ~\cite{koralz} program. A comparison of the cross-section predicted 
by KORALZ 4.02 with that predicted by NUNUGPV~\cite{nunugpv} and 
KK 4.19~\cite{kk2f} showed agreement at the percent level. This difference is negligible 
with respect to the statistical and systematic uncertainties in the 
present measurement. 
\par Simulated events for 
the irreducible contribution from $\nu\bar{\nu}\gamma$ production 
and other SM backgrounds were
generated at the different centre-of-mass energies and passed through the 
full DELPHI simulation and reconstruction chain~\cite{delphi}.


\vspace{0.4cm}
\begin{center}
\begin{tabular}{|c|c|c|c|} \hline
   & N$_{observed}$ & N$_{e^+{e^-} \rightarrow \nu{\bar{\nu}}({\gamma)}}$ & 
     N$_{other~SM~background}$  \\ \hline
FEMC & 705 & 626$\pm$3 & 49.1 \\ \hline
HPC  & 498 & 540$\pm$4 & 0.6  \\ \hline
\end{tabular} \\
\vspace{0.3cm}
Table 1: The number of selected and expected single-photon events. 
\end{center}
\vspace{0.4cm}

\par
\begin{figure}[ht!]
\vspace*{-0.5cm}
\begin{center}
\mbox{\epsfig{figure=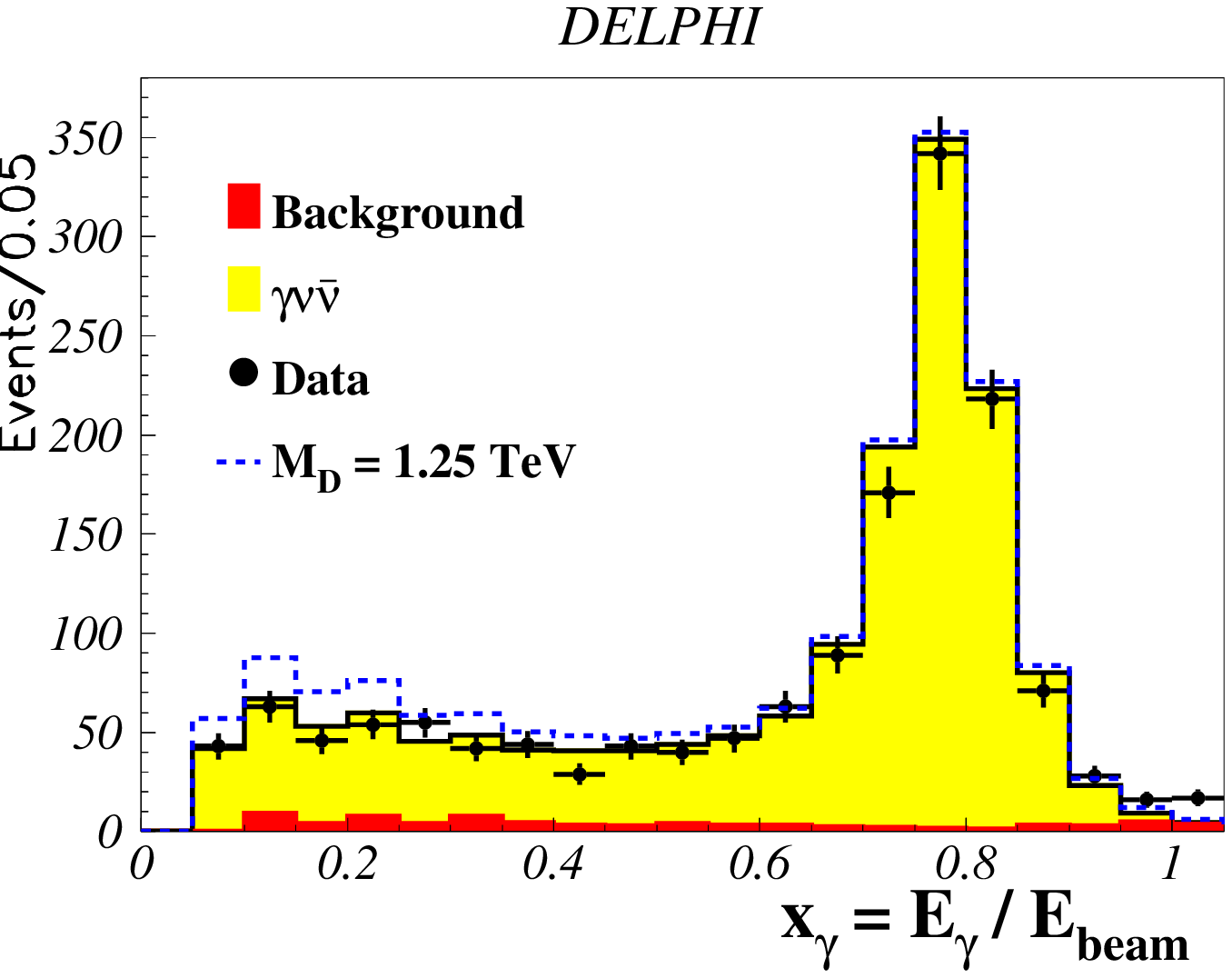,width=9.0cm}}
\end{center}
\caption[]{$x_\gamma$ of selected single photons.
The light shaded area is the expected distribution from \eenngg
and the dark shaded area 
is the total background from other sources. Indicated in the plot is also 
the signal expected from \eeGg
for $n$=1 and $M_D$=1.25 TeV. 
}
\label{fig:xgamma}
\end{figure}

Figure~\ref{fig:xgamma} shows the $x_\gamma$ distribution of all preselected 
single-photon events.
As discussed in the previous paper~\cite{de}, only single photon events 
in the HPC and FEMC were used for the subsequent analysis, since the 
$E_\gamma$ cuts in the STIC, needed to reduce the radiative Bhabha background, 
reject a large part of the ED signal even in the case $n$=1. 

\par Table 1 shows the total number of observed and expected events in the
HPC and FEMC.
The numbers are
integrated over the LEP energies from 180 to 209 GeV and correspond
to an overall luminosity of $\sim$650 $\mathrm {pb^{-1}}$.

\par  
A likelihood ratio method was used to select the final sample of 
single-photon events. The photon energy was used as the final 
discriminating variable and two likelihood functions ($f_S(E_\gamma$) 
and $f_B(E_\gamma$)) were produced from the normalized photon
energy distributions of the 
expected ED 
and SM background events, 
after passing through the same selection criteria. The likelihood ratio 
function was defined as ${\cal L}_R = f_S(E_\gamma)/f_B(E_\gamma$) 
where an event with ${\cal L}_R$ $>$ ${\cal L}_R^{CUT}$ was selected as a 
candidate event. The value of ${\cal L}_R^{CUT}$ was optimized on simulated events to give 
the minimum signal cross-section excluded at 95$\%$ CL in the absence 
of a signal:

\begin{equation}
\sigma^{min}({\cal L}_R^{CUT}) = {N_{95}^{min}({\cal L}_R^{CUT}) 
\over \epsilon^{max}({\cal L}_R^{CUT}) \times L},
\label{eq:sigmamin}
\end{equation}

\noindent where $N_{95}^{min}$ is the upper limit on the number of signal events 
at 95$\%$ CL, $\epsilon^{max}$ is the efficiency for the signal and $L$ is 
the integrated luminosity. This method optimises the background 
suppression for a given signal efficiency~\cite{anderson}.
The signal shape, $f_S(E_\gamma)$, is the only difference with respect to 
the previous analysis~\cite{de}. 
\par
The data collected at different centre-of-mass energies were analysed 
separately and different analyses were made depending on the 
electromagnetic calorimeter in which the photon was recorded. The 
final experimental limit was obtained using a Bayesian multi-channel 
method~\cite{obraztsov} which combined the results of 20 
analyses, the data for the two calorimeters being grouped
into 10 datasets between 180 and 209 GeV centre-of-mass 
energy. The method takes into account all the available information
(such as the fraction of the signal and the average background in 
each subdetector and in each data subsample), and this makes it 
possible to calculate optimum limits.   

\section{Limit on the production of gravitons}

\par The differential cross-section for \eeGg
has been calculated in~\cite{{grw},{gps}} and is given by 

\begin{equation}
 {{d^2}\sigma \over d{x_\gamma}dcos{\theta_\gamma}} = 
  {\alpha \over 32s} {\pi^{n \over 2} \over \Gamma({n \over 2})} 
  \left({\sqrt{s} \over M_D}\right)^{n+2} f({x_\gamma},cos{\theta_\gamma})  
\label{eq:dsdxdcost}
\end{equation}

with 

\begin{equation}
 f(x,y) = {2(1-x)^{{n \over 2}-1} \over x(1-y^2)}[(2-x^2)(1-x+x^2) \\
  - 3{y^2}{x^2}(1-x) - {y^4}{x^4}].
\label{eq:fxy}
\end{equation}

Initial state radiation can produce additional photons that would cause 
a signal event to be rejected in a single-photon analysis. The expected 
signal cross-section has therefore been corrected with a radiator 
approximation method~\cite{nicrosini}. 
\par
\begin{figure}[ht!]
\vspace*{-0.5cm}
\begin{center}
\mbox{\epsfig{figure=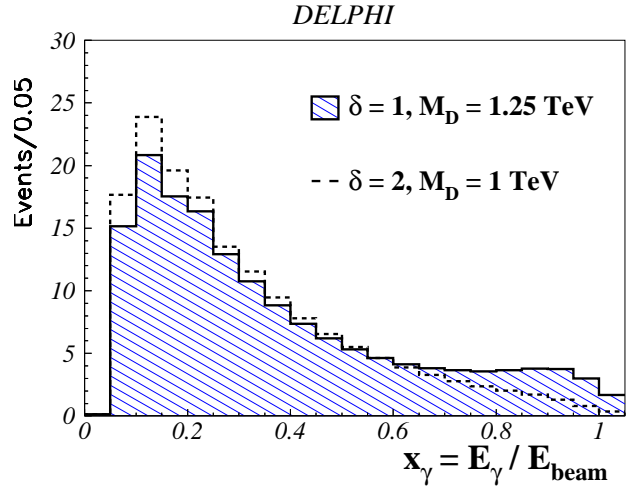,width=9.0cm}}
\end{center}
\caption[]{$x_\gamma$ of expected single photons in the HPC and FEMC 
from \eeGg
with $n$=1, $M_D$=1.25 TeV and $n$=2, $M_D$=1 TeV, 
corrected for calorimeter efficiency and resolution. 
MC expectations are normalized to the luminosity of the 
combined data set in Fig. 1.
}
\label{fig:esmeared}
\end{figure}

 For $n>$1 the differential distribution, eq.~\ref{eq:fxy}, is peaked at 
small $E_\gamma$ and $\theta_\gamma$, for $n$=1 instead a singularity is 
present at $x_\gamma$=1, which makes the distribution qualitatively 
different from the others. For instance the ratio of the cross-sections,
eq.~\ref{eq:dsdxdcost} and eq.~\ref{eq:fxy},  
for $n$=1 and $n$=2 is independent of $\theta_\gamma$, and increases 
from $\sim$1 at small $x_\gamma$ to $\sim$30 at $x_\gamma$=0.95 
for $M_D$=1 TeV and $\sqrt{s}$=208 GeV. 
In order to take into account detector effects,
the theoretical ED cross-section has been corrected for efficiency and 
energy resolution in the calorimeters, using a parameterization 
developed in the $\nu\bar{\nu}\gamma$ analysis. The theoretical energy 
distributions for $n$=1 and 2 smeared in the HPC and FEMC 
are shown in Fig.~\ref{fig:esmeared}. 
\par As can be seen in Fig.~\ref{fig:xgamma}, the single photon 
data measured by DELPHI were well compatible with expectations
from SM processes and no evidence for graviton production was found. 
\par All DELPHI data with $\sqrt{s}>$180 GeV were used and 
a dedicated selection for 
each bin in $\sqrt{s}$ was made 
as described in the 
previous section. These limits were combined to give a 95$\%$ CL 
cross-section limit 
for one extra dimension of 0.171 $\mathrm {pb}$ at 208 GeV, 
with an expected limit of 0.166 $\mathrm {pb}$. 
In terms of the parameter $p = (1/M_{D})^3$, which is proportional to the n=1
signal cross section, the combined log-likelihood function of the Bayesian 
formula was practically parabolic. 
$p$ is estimated to be ($0.009 \pm 0.098$) TeV$^{-3}$ and is therefore 
consistent with zero.  
The obtained limit on the fundamental mass scale is $M_D>$1.69 
TeV at 95$\%$ CL (with 1.71 TeV expected limit) in the $n$=1 analysis. 
As a comparison, the 
cross-section limits in the previous analysis for $n$=2-6 varied 
between 0.14 and 0.18 $\mathrm {pb}$, and the obtained limits for  $M_D$ 
between 1.31 TeV ($n$=2) and 0.58 TeV ($n$=6).
The same systematic errors were considered as in the
previous analysis~\cite{de}, namely trigger and identification efficiency, calorimeter energy scale and background, and the systematic error on the $M_D$
limit in the n=1 analysis was estimated to be less than 4\%.

\section{Conclusions}

We have re-analysed single-photon events detected with DELPHI 
at LEP2 
during 1997-2000 at centre-of-mass energies between 180 and 209 GeV 
to study graviton production with $n$=1 large extra dimensions, 
motivated by the model of Giudice, Plehn and Strumia~\cite{gps}. 
Since the measured single-photon cross-sections are in agreement 
with the expectations from the SM process \eenngg, the absence 
of an excess of events has been used to set a limit of 1.69 TeV 
at 95$\%$ CL on the fundamental mass scale for $n$=1 ED.



\clearpage



\end{document}